\documentclass[prd,superscriptaddress,floatfix,showpacs,10pt,letterpaper]{revtex4}
\usepackage{graphicx} \usepackage{subfigure} \usepackage{amssymb}
\usepackage{verbatim} \usepackage{amsmath} \usepackage{amscd}
\usepackage{amssymb} %% Load more math symbols
\usepackage{epsfig}     %% Load more math symbols

\begin{document}

\title{Measures for a Multidimensional Multiverse}

\author{Hyeyoun Chung}  
\email{hyeyoun@physics.harvard.edu.}
\affiliation{Center for the Fundamental Laws of Nature, Harvard University,\\ 17 Oxford St., Cambridge, MA 02138, USA}

\date{\today} \begin{abstract} We explore the phenomenological implications of generalizing the causal patch and fat geodesic measures to a multidimensional multiverse, where the vacua can have differing numbers of large dimensions. We consider a simple model in which the vacua are nucleated from a $D$-dimensional parent spacetime through dynamical compactification of the extra dimensions, and compute the geometric contribution to the probability distribution of observations within the multiverse for each measure. We then study how the shape of this probability distribution depends on the timescales for the existence of observers, for vacuum domination, and for curvature domination ($t_{obs}, t_{\Lambda},$ and $t_c$, respectively.) In this work we restrict ourselves to bubbles with positive cosmological constant, $\Lambda$. We find that in the case of the causal patch cutoff, when the bubble universes have $p+1$ large spatial dimensions with $p \geq 2$, the shape of the probability distribution is such that we obtain the coincidence of timescales $t_{obs} \sim t_{\Lambda} \sim t_c$. Moreover, the size of the cosmological constant is related to the size of the landscape. However, the exact shape of the probability distribution is different in the case $p = 2$, compared to $p \geq 3$. In the case of the fat geodesic measure, the result is even more robust: the shape of the probability distribution is the same for all $p \geq 2$, and we once again obtain the coincidence $t_{obs} \sim t_{\Lambda} \sim t_c$. These results require only very mild conditions on the prior probability of the distribution of vacua in the landscape. Our work shows that the observed double coincidence of timescales is a robust prediction even when the multiverse is generalized to be multidimensional; that this coincidence is not a consequence of our particular universe being (3+1)-dimensional; and that this observable cannot be used to preferentially select one measure over another in a multidimensional multiverse.
\end{abstract}

\pacs{} \maketitle %\narrowtext

\section{Introduction}\label{sec-Intro}

The measure problem is concerned with the issue of making predictions in an eternally inflating multiverse. As every single event that can happen, happens infinitely many times in such a multiverse\cite{Guth}, it is necessary to introduce a regulator of some sort in order to define relative probabilities\cite{FreivogelReview}. This regulator is called a \textit{measure}. Several reasonable measures have been put forward, which give different phenomenological predictions for relative probabilities of events occurring in the multiverse\cite{CAHPlus, ScaleFactor, BoussoBoundary, ScaleFactorCC, CPBousso, CPBousso1}. At first sight the existence of an eternally inflating multiverse might appear to be inconvenient and disconcerting, especially since the calculation of probabilities within such a multiverse is sensitively dependent on the choice of measure, and there is no way at present of establishing a particular choice of measure as the correct one. However, Weinberg's prediction of the cosmological constant\cite{Weinberg} showed that the existence of a multiverse can be used to explain the smallness of the observed value of $\Lambda$, and the fact that $\Lambda \neq 0$, by using anthropic arguments. This result has become a compelling reason for studying the calculation of probabilities within a multiverse, in the hopes that we can find measures that will explain some of the observations that we make in our own universe, such as the values we measure for certain physical constants, or the coincidence of the timescales of vacuum domination, and the timescale at which observations are made\cite{GarrigaProb, Feldstein, QCatastrophe, Graesser, GarrigaPrediction, ProbBoussoPolchinski, FreivogelObsvConsq, Vilenkin, Salem, BoussoPhenom, BoussoPhenom1}. If we assume the existence of an infinite and eternally inflating multiverse, where these constants take different values in each of the bubble universes, then it could be that we are highly likely to find ourselves in a universe where the parameters of physical theories take the values that we observe.

One such observable that is a feature of our own universe is the coincidence of three separate timescales: for the existence of observers, for vacuum domination, and for curvature domination ($t_{obs}, t_{\Lambda},$ and $t_c$ respectively), so that $t_{obs} \sim t_{\Lambda} \sim t_c$. Recent work by Bousso et al. showed that this coincidence can be predicted from a variety of different measures using solely geometrical arguments\cite{BoussoPhenom, BoussoPhenom1}. This analysis assumed that all of the vacua in the multiverse under consideration are $(3+1)$-dimensional; however, there is no guarantee that this is the case. In fact, it is easy to envision a fundamental theory which allows for the nucleation of vacua with different numbers of large dimensions, with the extra dimensions being compactified on an internal manifold: the string theory landscape, for example, is expected to contain a large landscape of vacua with different numbers of dimensions.

This point naturally raises an interesting question: is the coincidence of timescales $t_{obs} \sim t_{\Lambda} \sim t_c$ still predictable using a measure when the multiverse is multidimensional? And is the double coincidence that we observe, somehow a consequence of our universe being (3+1)-dimensional? We address this problem in this paper. First we consider two local measures that have been found to be phenomenologically satisfactory in the case of (3+1)-dimensional vacua--the causal patch measure\cite{CPBousso, CPBousso1}, and the fat geodesic measure\cite{FreivogelReview, ScaleFactor, ScaleFactorCC}--and generalize them straightforwardly to the case of a multi-dimensional multiverse. We then study how these measures can be used to predict the timescales of vacuum domination, curvature domination, and observation ($t_\Lambda, t_c$, and $t_{obs}$) in the various vacua, by applying the measures to a specific, yet sufficiently generic model of a multiverse with vacua of differing numbers of large dimensions. Note that in this work, we restrict ourselves to analyzing vacua with positive cosmological constant $\Lambda$.

We follow the approach of Bousso et al., who investigated the ways in which the geometry of various measures could affect their phenomenological predictions\cite{BoussoPhenom, BoussoPhenom1}. The general method is to determine the probability distribution of observations in the multiverse over the three timescales $t_\Lambda, t_c$, and $t_{obs}$, by factoring the distribution into a part corresponding to the prior probability of the formation of bubbles characterized by the parameters ($t_\Lambda, t_c$), a part corresponding to the density of observers per unit mass per logarithmic time interval, averaged over the different types of bubbles, and a part corresponding to the mass inside the cutoff. By arguing that the first two factors have a constrained form that we can determine through logical reasoning, and that the leading contribution to the probability distribution comes from the third and last term, which can be explicitly calculated, we can then study important features of the probability distribution.

There are several ways in which the probability distribution of observations calculated in this manner could display some qualitative differences in the multi-dimensional case, as compared to the (3+1)-dimensional case. If we assume that the vacuum bubbles in the multi-dimensional case are open FRW universes with the extra dimensions compactified on a sphere, then the evolution of the FRW scale factor inside the bubble, and thus the mass contained inside the cutoff, will depend on the number of large dimensions. The parameter $t_\Lambda$, which is related to the effective cosmological constant inside the bubble, is also related to the number of large dimensions. Thus the probability distribution could change depending on the number of large dimensions there are in the bubbles that we are considering. Moreover, the probability distribution could depend, not just on the number of large dimensions in the bubble, but the number of large dimensions relative to the total dimension $D$ of the original higher-dimensional space. In practice we find that the shape of the probability distribution is largely determined by just one factor: the number of large dimensions in the bubble. For the causal patch measure, the results can be classed into three main groups: the case of 2 large spatial dimensions, 3 large spatial dimensions, and all spatial dimensions larger than three. However, although the probability distributions differ in all three cases, for 3 spatial dimensions or larger, we predict the coincidence $t_{obs} \sim t_{\Lambda} \sim t_c$. For the fat geodesic measure, the results can be classed into two groups: the case of 2 large spatial dimensions, and the case of 3 or more large spatial dimensions. Finally, for both the causal patch measure and the fat geodesic measure, the smallness of the cosmological constant is related to the number of vacua in the multiverse.

This paper is structured as follows. In Section \ref{sec-Measures}, we describe the causal patch and fat geodesic measures and outline how they can be generalized to the case of a multiverse with multi-dimensional vacua. In Section \ref{sec-MultiModel}, we describe the specific model that we will use for such a multiverse, so that we can carry out concrete calculations using the measures. In Section \ref{sec-Phenom} we explore the phenomenological properties of the measures by applying them to this multiverse model. We conclude in Section \ref{sec-Conclusion}.

\section{The Measures and their Generalizations to a Multi-dimensional Multiverse}\label{sec-Measures}

We can immediately generalize two well-known measures to the case of a multi-dimensional multiverse: the \textit{causal patch} measure\cite{CPBousso, CPBousso1}, and the \textit{fat geodesic} measure\cite{FreivogelReview, ScaleFactor, ScaleFactorCC}. Both are local measures, which define the relative probabilities of different events by only counting events in a finite neighborhood of a single inextendible timelike geodesic in the multiverse (where the neighborhood is defined differently for each measure), and then taking an average over initial conditions and possible decoherent histories for the geodesic. If $N_I$ is the number of times that outcome $I$ occurs within the specified neighborhood of the geodesic, and $\langle N_I \rangle$ indicates the expectation value after averaging over initial conditions and decoherent histories, then the relative probability of outcomes $I$ and $J$ are given by:
\begin{align}
\frac{p_I}{p_J} &= \frac{\langle N_I \rangle}{\langle N_J \rangle}
\end{align}
We can find the expectation value $\langle N_I \rangle$ by constructing an ensemble of geodesics, and then taking the ensemble average. This is done by selecting an initial spacelike hypersurface $\Sigma_0$, and then constructing a geodesic orthogonal to that hypersurface. If we then take $Z$ identical copies of $\Sigma_0$ and choose the same starting point for the geodesic in each copy, the resulting $Z$ geodesics, and their corresponding neighborhoods, represent different decoherent histories for the multiverse. In order to account for different initial conditions, we can then take a weighted average over different initial surfaces $\Sigma_0$, which correspond to different initial vacua.

The \textbf{causal patch measure} is a local measure for which the local neighborhood of the geodesic is taken to be the causal patch of the endpoint of the geodesic. The \textbf{fat geodesic measure} is a local measure for which the local neighborhood of the geodesic is a fixed infinitesimal orthogonal cross-sectional volume $dV$, which we take to be spherical. Both of these measures can be directly generalized to the case of a multidimensional multiverse with no change in their definitions.

\section{Determining the Probability Distribution of Observations}\label{sec-Outline}

Here we outline the general approach taken in \cite{BoussoPhenom, BoussoPhenom1} to determine the essential features of the probability distribution of observations in the multiverse. This probability distribution over the time of existence of observers, the time of curvature domination, and the time of vacuum domination, $(\log t_{obs}, \log t_c, \log t_\Lambda)$, takes the form:
\begin{align}\label{eq-ProbDens}
\frac{d^3p}{d\log t_{obs}d\log t_\Lambda d\log t_c} &= \frac{d^2\tilde{p}}{d\log t_\Lambda d\log t_c}\times M (\log t_{obs},\log t_c, \log t_\Lambda) \times\alpha(\log t_{obs},\log t_c, \log t_\Lambda).
\end{align}
The first factor is the \textit{prior probability distribution}, which corresponds to the probability of nucleating a bubble with parameters $(\log t_c, \log t_\Lambda)$ inside the cutoff region. The second and third factors combined give the probability density for observations within a given bubble. This probability density is further divided into the mass $M$ inside the cutoff region, and $\alpha$, which is the average number of observations per unit mass per time.

It is possible to calculate $M$ explicitly for each cutoff. We carry out this calculation in Section \ref{sec-Phenom}, for both the causal patch and fat geodesic cutoffs. We also argue that $\alpha$ can be written as a function purely of $\log t_{obs}$, which requires only weak assumptions\cite{BoussoPhenom, BoussoPhenom1}. Furthermore, we expand the form of the prior probability density $\tilde{p}$ as:
\begin{align}
\frac{d^2\tilde{p}}{d\log t_\Lambda d\log t_c} \sim t_\Lambda^{-2} g(\log t_c)
\end{align}
for some function $g(\log t_c)$, as we are considering small values of $\Lambda \sim t_\Lambda^{-2}$ and can thus Taylor expand in $\Lambda$\cite{Weinberg}.

%Probability distribution not over t_\Lambda, but other things? p? q? Q? \Lambda_D?

%In bubbles with positive cosmological constant, the calculable quantity $M$ so strongly suppresses the probability in other regimes that in many cases we only need to know the form of $\alpha$ in the regime where observers live before vacuum energy or curvature become important, $t_{obs} < t_\Lambda, t_c$. Under very weak assumptions, $\alpha$ must be independent of $t_\Lambda$ and $t_c$ in this regime. This is because neither curvature nor vacuum energy play a dynamical role before observers form, so that neither can affect the number of observers per unit mass. Thus, for positive cosmological constant $\alpha$ is a function only of $t_{obs}$ in the only regime of interest.

\section{A model for a multi-dimensional multiverse}\label{sec-MultiModel}

In order to explicitly explore the phenomenological predictions of these generalized measures, we must specify a concrete model for a multiverse where the vacua can have different numbers of large dimensions, and apply the measures to this model. We would like the model to be fairly generic, so that we can identify the qualitative properties of the measures that result from extending them to multi-dimensional models, without our results being too much influenced by the particular model we have chosen.

Starting with a fundamental theory in $D$-dimensions, we would like to consider a model that allows for the dynamic nucleation of vacua with varying numbers of large dimensions, while the remaining spatial dimensions are compactified. In general, the Lagrangian for the fundamental theory will have the form
\begin{align}
S_D &= \int d^D x \sqrt{-\tilde{g}^{(D)}} \left [ M_D^{D-2} \tilde{\mathcal{R}} + \mathcal{L}(\psi) + \hat{\mathcal{L}}(\tilde{\mathcal{R}})\right ],
\end{align}
where $\tilde{g}^{(D)}$ is the metric on the full $D$-dimensional spacetime, $M_D$ is the $D$-dimensional Planck mass, $\tilde{\mathcal{R}}$ is the Ricci scalar, and $\mathcal{L}(\psi)$ gives the contribution of matter sources. The term $\hat{\mathcal{L}}(\tilde{\mathcal{R}})$ represents possible curvature corrections to the Einstein-Hilbert Lagrangian. In this work we assume that there are no terms that mix $\psi$ and $\tilde{\mathcal{R}}$.

Theories of this form were studied by Giddings in \cite{Giddings}, where the metric was assumed to have the form
\begin{align}
ds^2 = e^{2A(y)}ds_{4}^2 + R^2(x)g_{mn}(y) dy^m dy^n,
\end{align}
where $ds_{4}^2$ is the metric for (3+1)-dimensional de Sitter space. There are four large dimensions, and $D-4$ compact dimensions. A radial dilaton field $R(x) = e^{\phi(x)}$, that depends only on the coordinates of the large dimensions, encodes the size of the compact dimensions. By assuming that $R(x)$ varies slowly on scales of order the compactification size, the equations for the matter fields can be solved to give $\psi = \psi_0$. Substituting these solutions into the full action, and then integrating over the compact dimensions, gives an effective potential for the radial dilaton. The dimensionally reduced action has the form:
\begin{align}
S &= \int d^4 x \sqrt{-g_4} \left [ \mathcal{R}_4 - \frac{1}{2} (D-4)(D-2) (\nabla \phi)^2 + V(\phi) \right ]
\end{align}
where $V(\phi)$ is the effective potential for $\phi(x)$. Thus the dimensionally reduced theory has the form of Einstein gravity in $(3+1)$ dimensions coupled to a scalar field, the radial dilaton, whose dynamics are governed by an effective potential $V(\phi)$.

If this potential has stable or metastable local minima, then the nucleation of vacua with different numbers of large dimensions is possible in this theory. For example, if the potential has a local minimum such that the value of the potential at that minimum is positive, then this corresponds to a positive cosmological constant in the dimensionally reduced theory. The local minimum is a stable solution for the dilaton field, and so represents a de Sitter vacuum in $(3+1)$ dimensions where the extra dimensions are compactified on a $D-4$-dimensional sphere. If the value of the potential at the local minimum is negative, then the minimum is an AdS vacuum in $(3+1)$ dimensions; if the value of the potential is zero, then the minimum is a Minkowski vacuum. It is also possible for the effective potential to have several local minima and thus allow for several possible vacua. It is possible that some or all of these vacua are metastable; however, if the decay rates of these vacua are sufficiently low, then the metastable vacua correspond to valid, viable universes within the multiverse that could exist for long enough for observers to evolve.

\begin{figure}\includegraphics[width=250pt]{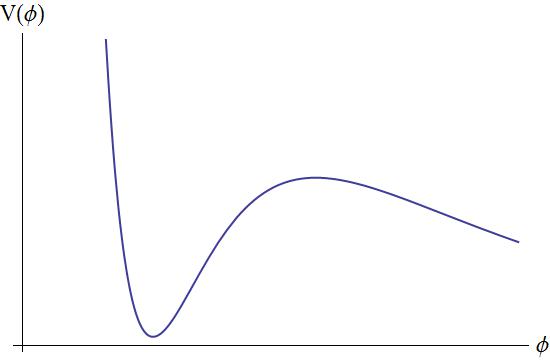}
\caption{An effective potential for the radial dilaton field $\phi(x)$ that has a metastable minimum corresponding to a vacuum with a positive effective cosmological constant.}
\label{fig-Potential}
\end{figure}

The nucleation of a $(3+1)$-dimensional vacuum in this theory can occur through the process of \textit{compactification}, which is exactly analogous to the nucleation of bubbles via Coleman-DeLuccia or Hawking-Moss instantons. In the Coleman-DeLuccia instanton, we consider a scalar field coupled to Einstein gravity that is in its false vacuum state throughout the whole of space. This corresponds to the scalar field taking its value at a local minimum of its potential. A vacuum of true vacuum can nucleate within the false vacuum, if the scalar field tunnels through the potential barrier to the global minimum of the potential. In the Hawking-Moss instanton, a bubble is nucleated in which the scalar field has tunneled to its value at a local \textit{maximum} of the potential, and then rolls down into the global minimum, so that the bubble becomes a bubble of true vacuum. If we think of the radion field $\phi(x)$ as a scalar field coupled to gravity in the $(3+1)$ large dimensions, then bubbles of vacua can nucleate in which $\phi(x)$ takes its values at local minima of the potential, which corresponds to the extra dimensions being compactified. It is also possible for vacua to be formed through \textit{decompactification} transitions, in which bubbles of vacua nucleate through $\phi(x)$ tunneling through the potential barrier in the other direction, from some finite value to $\phi \to \infty$, so that the extra dimensions go from being compact to large.

Giddings considers various possible matter sources for the dilaton potential, including fluxes in the compact dimensions, and branes wrapped around the compact dimensions. In both cases, he finds that the sources contribute terms to the effective potential that are exponentials of the dilaton field $\phi(x)$. Thus we can generically assume that the effective potential for $\phi(x)$ is a sum of exponentials of $\phi(x)$, independently of the particular model under consideration.

One explicit model of this form is that discussed by Carroll, Johnson, and Randall in \cite{Randall} and by Blanco-Pillado, Schwartz-Perlov, and Vilenkin in \cite{Vilenkin1}, which considers a fundamental theory of Einstein gravity in $D$ dimensions with a cosmological constant, coupled to one or more $q$-form field strengths. The action for the higher-dimensional theory is:
\begin{align}
S_D &= \frac{1}{2}\int d^D x \sqrt{-\tilde{g}^{(D)}}\left ( \tilde{\mathcal{R}}^{(D)} - 2\Lambda_D - \frac{1}{2q!}\tilde{F}_q^2\right ),
\end{align}
where we have used $\Lambda_D$ to denote the $D$-dimensional cosmological constant in order to distinguish it from the effective cosmological constant inside the FRW bubble universes that we will describe later. Assuming spherical symmetry in the compact dimensions, we can write the metric in the form
\begin{align}
ds^2 = \tilde{g}_{\mu\nu}^{p+2}(\textbf{x}) dx^\mu dx^\nu + R^2(\textbf{x})d\Omega_q^2,
\end{align}
where we have decomposed the metric into $(p+2)$ large dimensions and $q$ compact dimensions, and the entire space has $D = p+2+q$ dimensions. The radion field $R(x)$ and the $p+2$-dimensional metric $\tilde{g}_{\mu\nu}^{p+2}$ are functions of the $p+2$-dimensional coordinates $x$. The magnetic $q$-form field strengths solving Maxwell's equations and respecting the $q$-dimensional spherical symmetry are given by
\begin{align}
\tilde{F}_q &= Q\sin^{q-1}\theta_1\dots\sin\theta_{q-1}d\theta_1\wedge \dots \wedge d\theta_q
\end{align}
We consider solutions with only a single fixed $q$, so that multiple $q$-form charges are not turned on simultaneously. We can now integrate over the $q$ compact dimensions to obtain the dimensionally reduced theory in $p+2$ dimensions. We also perform a conformal transformation in order to express our results in the $p+2$-dimensional Einstein frame, so that we can view the theory as Einstein gravity coupled to a scalar field. The conformal transformation is
\begin{align}
g_{\mu\nu} &= R^{2\frac{q}{p}}\tilde{g}_{\mu\nu}
\end{align}
and the dimensionally reduced action in the Einstein frame is
\begin{align}
S_{p+2} = \int d^{p+2}x \sqrt{-g} \left [ \frac{1}{2}\mathcal{R} - \frac{1}{2}\frac{q(p+q)}{pR^2}g^{\mu\nu}\partial_\mu R \partial_\nu R - V(R)\right ],
\end{align}
where $V(R)$ is an effective potential for the radion field $R(x)$. We can define a canonically normalized radion field $\phi$ by making the change of variables
\begin{align}
R &= \exp \left [\sqrt{\frac{p}{q(p+q)}}\frac{\phi}{M_{p+2}}\right]
\end{align}
In terms of $\phi$, the dimensionally reduced action is
\begin{align}
S_{p+2} = \int d^{p+2}x \sqrt{-g} \left [ \frac{1}{2}\mathcal{R} - \frac{1}{2}g^{\mu\nu}\partial_\mu \phi \partial_\nu \phi - V(\phi)\right ]
\end{align}
where the effective potential $V(\phi)$ is
\begin{align}
V(\phi) &= \frac{1}{2}\left [ -q(q-1)\exp \left ( -2\sqrt{\frac{p+q}{pq}}\phi\right ) + 2\Lambda_D\exp\left ( -2\sqrt{\frac{q}{p(p+q)}}\phi \right ) + \frac{Q^2}{2}\exp\left ( -2(p+1)\sqrt{\frac{q}{p(p+q)}}\phi \right )\right ]
\end{align}
In general, we can have $J$ copies of a $q$-form, with $e_i$ being the gauge coupling for each copy, and $n_i$ being the number of units of fundamental charge, in which case $Q$ in the above formula is replaced with
\begin{align}
Q^2 \equiv \sum_{i=1}^J Q_i^2 = \sum_{i=1}^J e_i^2 n_i^2.
\end{align}
Thus, for a given number $D$ of dimensions in the fundamental theory, we have three independent parameters that we can vary: the number of compact dimensions $q$, the charge $Q$, and the higher-dimensional cosmological constant $\Lambda_D$. These parameters determine the form of the effective potential $V(\phi)$. It was shown in \cite{Randall} that the nucleation of vacua in this theory through decompactification transitions is exponentially suppressed relative to the probability of nucleating vacua through compactification transitions, so we will consider only the latter in this work. 

The possible forms $V(\phi)$ were studied in \cite{Randall}, and it was found that we require $\Lambda_D \neq 0$ for the existence of solutions that allow for the dynamical nucleation of bubbles with compactified dimensions. The results are qualitatively similar when $\Lambda_D < 0$ compared to when $\Lambda_D > 0$, so we will only consider the case $\Lambda_D > 0$. For certain values of $Q$, the effective potential can have a local minimum and a local minimum, showing that there can be $p+2$-dimensional metastable vacua. The value of the potential at the local minimum can be positive, zero, or negative, so that the metastable vacua can be de Sitter, Minkowski, or Anti de Sitter.

We would like to further generalize this model to allow for the presence of pressureless macroscopic matter, so that we can have regions of matter, curvature, and vacuum domination inside the bubbles nucleated in the multiverse. This can be done quite easily by using the Einstein equations in the $p+2$-dimensional Einstein frame:
\begin{align}
R_{\mu\nu} - \frac{1}{2}g_{\mu\nu}R = T_{\mu\nu},
\end{align}
where we have chosen units so that $8\pi G = 1$. If we compute the Ricci tensor and Ricci scalar for the $p+2$-dimensional metric $g_{\mu\nu}$, then on the right-hand side we can substitute the energy-momentum tensor $T_{\mu\nu}$ by summing the energy-momentum tensor for the scalar field $\phi$, and the energy-momentum tensor for pressureless matter, which has the form
\begin{align}
T_{\mathrm{matter}}^{\mu\nu} &= \rho U^{\mu}U^\nu \quad\quad\mbox{for}\qquad \mu = \nu = 0\\
T_{\mathrm{matter}}^{\mu\nu} &= 0 \quad\quad\mbox{otherwise.},
\end{align}
for some constant density $\rho$, where the density is calculated with respect to the large dimensions (as we are considering macroscopic matter.)

\section{Open FRW Universes with Compactified Dimensions}\label{sec-FRW}

In order to compute the mass contained within the cutoff region, the calculation in \cite{BoussoPhenom, BoussoPhenom1} proceeds by assuming that every bubble in the multiverse is an open FRW universe, as is usually formed by Coleman-deLuccia tunneling. The metric has the form:
\begin{align}
ds^2 = -dt^2 + a(t)^2(d\chi^2 + \sinh^2\chi d\Omega^2)
\end{align}
The Friedmann equations for $a(t)$ have the form
\begin{align}
\left (\frac{\dot{a}}{a}\right )^2 &= \frac{t_c}{a^3} + \frac{1}{a^2} \pm \frac{1}{t_\Lambda^2}
\end{align}
and can be solved for $a(t)$ piecewise, in terms of $t_\Lambda, t_c$, by ignoring all but one term on the right-hand side for each region in time. The nature of the solution depends on whether the cosmological constant $\Lambda$ (related to $t_\Lambda = \sqrt{\frac{p(p+1)}{2|\Lambda|}}$) is positive or negative, and whether $t_c \ggg t_\Lambda$ or $t_c \lll t_\Lambda$.

%%%%%%%%%%%%%%%

We can adapt this analysis to our model (though in this work we only consider positive cosmological constant $\Lambda > 0$.) A bubble universe can nucleate in the background $D$-dimensional multiverse through a process analogous to Coleman-DeLuccia tunneling\cite{Randall}, with the radion field $\phi$ playing the role of the scalar field in the CdL instanton, as described in Section \ref{sec-MultiModel}. The spacetime after nucleation is given by a configuration analogous to a thick-walled CdL bubble. The region inside the bubble is given by an open FRW universe, where the metric in the $p+2$-dimensional Einstein frame has the form
\begin{align}
ds^2 = -dt^2 + a(t)^2(d\chi^2 + \sinh^2\chi d\Omega_p^2)
\end{align}
In order to match to the solution outside the bubble, the solution inside the bubble must satisfy the conditions $a = 0$, $\dot{\phi} = 0$ at $t=0$.

The equations of motion for $a(t)$ and $\phi(t)$ give us the multi-dimensional analogue of Friedmann equations, which take the form:
\begin{align}
\left (\frac{\dot{a}}{a}\right )^2 &=  \frac{t_c^{p-1}}{a^{p+1}} + \frac{1}{a^2} \pm \frac{1}{t_\Lambda^2} + \frac{2}{p(p+1)}\left (\frac{\dot{\phi}^2}{2} + V(\phi) - V_0 \right )\\
\ddot{\phi} + (p+1)\frac{\dot{a}}{a}\dot{\phi} &= -V'(\phi)
\end{align}
where $V(\phi)$ is the potential for $\phi$, and $V_0$ is the constant part of $V(\phi)$ (which gives the cosmological constant term $\frac{1}{t_\Lambda^2}$.) The term $\frac{t_c^{p-1}}{a^{p+1}}$ on the righthand side of the Friedmann equation for $a(t)$ corresponds to the contribution of non-dynamical, pressureless matter. We assume that the matter is macroscopic, so that it scales as $\sim \frac{1}{a^{p+1}}$. It would also be possible to analyze the effect of microscopic matter that is at the scale of the compact dimensions, by introducing a term $\sim \frac{1}{a^{p+1}R^q}$, where $R$ is the radion field. Preliminary computations indicates that this does not make a significant qualitative difference to our results, so we will consider only macroscopic matter in this paper and leave a more detailed analysis to a future work.

We cannot solve these equations analytically. However, we can solve them numerically. The potential $V(\phi)$ can be specified following the model given in \cite{Randall} (which, as shown in \cite{Giddings}, is a fairly generic potential that can be obtained by dimensionally reducing different fundamental theories, and so can be considered to be largely model-independent.) The potential depends on the number of large spatial dimensions $(p+1)$, the number of compact dimensions $q$, the cosmological constant in the full $p+2+q$-dimensional theory $\Lambda$, and the sum of charges of the $q$-form fluxes in the compact dimensions, $Q$. By scanning over these parameters, we can find a large set of solutions for different values of $t_c$ and $t_\Lambda$, and see how these solutions vary with each of the parameters.

In general, a solution for a bubble is obtained through a Coleman-DeLuccia instanton for $\phi$, where $\phi$ tunnels through a potential barrier in $V(\phi)$ before emerging at some value $\phi_-$, and then rolls down into a local minimum of $V(\phi)$, then oscillates about that minimum before settling down to a constant value, to give an FRW universe with $q$ compact dimensions with $\phi$ at some finite value $\phi_{min}$. If the value of the potential at $\phi_{min}$ is positive then we obtain an asymptotically de Sitter bubble, and $a(t)$ increases monotonically from zero to infinity, and if the value of the potential at $\phi_{min}$ is positive then we obtain an asymptotically $AdS$ bubble, and $a(t)$ crunches back to zero. If the value of the potential at $\phi_{min}$ is zero, then we obtain an asymptotically $M_{p+2}$ bubble.

\begin{figure}\includegraphics[width=250pt]{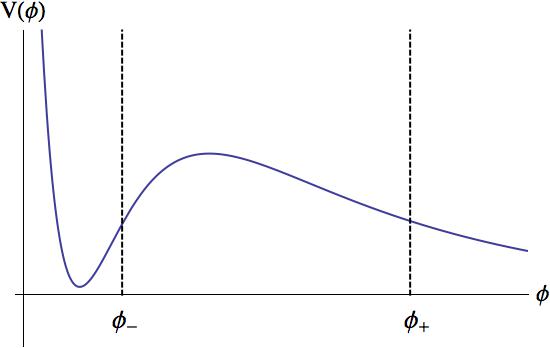}
\caption{An effective potential for the radial dilaton field $\phi(x)$ that has a metastable minimum corresponding to a vacuum with a positive effective cosmological constant. Compactification can occur dynamically when the field $\phi$ tunnels through the potential barrier between $\phi_+$ and $\phi_-$, and then rolls down into a local minimum of $V(\phi)$, which corresponds to a de Sitter universe in this case, as the value of $V(\phi)$ at this minimum is positive.}
\label{fig-Potential2}
\end{figure}

Thus, technically we would need to calculate the value $\phi_-$ in order to obtain the correct initial value of $\phi$ within the FRW bubble. However, we have verified numerically that starting with a fairly arbitrary initial value of $\phi_-$, between the local minimum and local maximum of the potential, does not significantly affect the final solutions $a(t), \phi(t)$. Therefore, as it is difficult to compute the CdL instanton precisely, in order to numerically solve the Friedmann equations we start with an arbitrary initial value for $\phi(t)$, chosen to be halfway between the values of $\phi$ at the local minimum and local maximum of the effective potential.

The work of Bousso et al. solves the equations analytically by ignoring the scalar field $\phi$ that drives the Coleman-deLuccia tunneling process (note that in general the field $\phi$ is not necessarily the radial dilaton field, but any scalar field that can lead to bubble nucleation.) We could do the same in this case, by considering only the time period after which the contribution of $\phi$ to the differential equations for $a(t)$ become subdominant, or conditioning on this being the case. Numerical analysis shows that in the case of $p=2$, where there are three large spatial dimensions (precisely the case studied in previous investigations of the phenomenology of measures) this gives the same results for the probability density of observations as directly solving the coupled equations for $a(t)$ and $\phi(t)$, and thus it is valid to ignore the evolution of $\phi(t)$. However, we find that when $p \neq 2$, taking into account the time-dependence of $\phi(t)$ leads to qualitatively different results compared to the analytical results obtained when $\phi(t)$ is ignored. Thus we proceed with the full numerical analysis. As we are interested in the general trends in the probability density of observations (i.e. whether the probability density favors larger or smaller values of $t_c$ and $t_\Lambda$), rather than the precise shape of the probability density function, we believe that specifying a particular model for the potential $V(\phi)$ will not detract from the general applicability of our results, especially since (as we have already noted) $V(\phi)$ itself is fairly generic and takes similar functional forms when derived from several different fundamental models. Thus we are taking into account not so much the effect of the precise shape of $V(\phi)$, but rather the existence of a time-dependent scalar field $\phi(t)$ whose evolution is related to the evolution of the scale factor $a(t)$. As $\phi(t)$ in this case is the radial dilaton that encodes the size of the compact dimensions, and we are specifically considering how the number of large vs. compactified dimensions affects the phenomenology of measures in a multiverse, it is appropriate that we consider the full coupled equations that describe the time-evolution of both $\phi(t)$ and $a(t)$. Moreover, this allows us to consider a wider range of possible $t_c$ and $t_\Lambda$, without having to assume that $\phi$ is sub-dominant in the region that we are considering.

\section{The Phenomenology of the Measures}\label{sec-Phenom}

We can now proceed to calculate the mass contained within the cutoff in a typical bubble, and thus find part of the contribution to the probability distribution of observations in the multiverse.

\subsection{The causal patch measure}\label{sec-CP1}

For the causal patch measure, we first use boost symmetries to say that the center of the causal patch is at the origin of a particular bubble, given by $\chi = 0$. Then for each time $t$ inside the bubble, the maximum value of $\chi$ contained inside the cutoff region is given by:
\begin{align}
\chi_{CP}(t) = \int_t^{t_f} \frac{1}{a(t')} dt',
\end{align}
as this is the range of $\chi$ contained in the past light-cone of a comoving geodesic that begins at the origin. As we are only considering $\Lambda > 0$, where $\Lambda$ is the effective cosmological constant inside the bubble (as opposed to the $D$-dimensional cosmological constant $\Lambda_D$), then we may take $t_f\to\infty$, by neglecting late-time decays inside the bubble.

We now want to calculate the mass of observers contained within the causal patch. Recall that we have assumed the observers are macroscopic, and thus scale as $\sim \frac{1}{a^{p+1}}$, i.e. the volume of the large dimensions in the spacetime, as explained in Section \ref{sec-FRW}. Thus the mass of observers is given by $M_{CP} = \rho a^{p+1}V_{CP} = t_c^{p-1} V_{CP}$, where $\rho$ is the matter density, and $V_{CP}$ is the macroscopic volume of the large dimensions inside the bubble available to observers at $t_{obs}$ (not the total volume inside the bubble, which we would need to compute if we were considering non-macroscopic matter and observers.) This quantity may be calculated by:
\begin{align}
V_{CP} = \mathcal{S}[\chi_{CP}(t_{obs})],
\end{align}
where $\mathcal{S}[\chi]$ is the comoving volume inside a sphere of radius $\chi$. If we were considering non-macroscopic observers that scale as $\sim \frac{1}{a^{p+1}R^q}$, where $R$ is the radion field, then we would have to compute the total volume inside the bubble, $V_{CP}^{tot}$. The mass inside the volume would be given by $M_{CP} = \rho a^{p+1}R^q V_{CP}^{tot} = t_c^{p-1} V_{CP}^{tot}$, as in this case we would have $t_c^{p-1} = \rho a^{p+1}R^q$.

The probability distribution of observations made in the landscape is then given by:
\begin{align}\label{eq-ProbDistCP}
\frac{d^3p}{d\log t_{obs}d\log t_\Lambda d\log t_c} &= \sim t_\Lambda^{-2} g(\log t_c)\times \alpha(\log t_{obs},\log t_c, \log t_\Lambda) \times M_{CP} (\log t_{obs},\log t_c, \log t_\Lambda) 
\end{align}
If we assume that the factors $g(\log t_{obs})$ and $\alpha(\log t_{obs}, \log t_\Lambda, \log t_c)$ do not provide the leading contributions to the probability distribution, but rather the factor $M(\log t_{obs}, \log t_\Lambda, \log t_c)$ does, then we can calculate the contribution of $M_{CP}$ to the probability density and analyze the resulting trends.

In order to calculate the probability distribution over the three variables $t_{obs}, t_\Lambda, t_c$, we considered a range of $p$ and $q$, and identified values of the parameters $Q$ and $\Lambda$ that would give a potential with a positive-valued local minimum, so that we could look at vacua with a positive effective cosmological constant. Scanning over different ranges of $Q$ and $\Lambda$ gave us a range of $t_\Lambda$ to work with. We solved the differential equations for $a(t)$ and $\phi(t)$ for a large range of $t_\Lambda$ and $t_c$ (where $t_c$ was input by hand), and then calculated $M_{CP}$ for a range of $t_{obs}$, choosing values for the three time scales so that they were well separated. We considered the following regions of parameter space:
\begin{align}
&\mbox{Region I}\qquad\qquad t_{obs} < t_c < t_\Lambda\\
&\mbox{Region II}\qquad\qquad t_c < t_{obs} < t_\Lambda\\
&\mbox{Region III}\qquad\qquad t_c < t_\Lambda < t_{obs}\\
&\mbox{Region IV}\qquad\qquad t_\Lambda < t_{obs}\\
&\mbox{Region V}\qquad\qquad t_{obs} < t_\Lambda\\
\end{align}
where we fixed $t_c \lll t_\Lambda$ in Regions I, II, and III, and $t_\Lambda \lll t_c$ in Regions IV and V. In the cases of Regions IV and V, we do not need to consider the size of $t_c$ relative to $t_{obs}$, as curvature domination never occurs when $t_\Lambda \lll t_c$. We then plotted the calculable parts of the probability distribution, namely $t_\Lambda^{-2}M_{CP}$, for each region and a variety of $(p,q,Q,\Lambda)$ to see if there were any discernible patterns in the results.

We found that the qualitative results are largely independent of the number of compact dimensions $q$ (regardless of the total number of dimensions $D = p+2+q$), and the values of $Q$ and $\Lambda$. The overall trends in the probability distribution are instead largely determined by the number $p+1$ of large dimensions in the nucleated bubbles. These results are discussed in Sections \ref{sec-Dimp2}-\ref{sec-Dimp1}.

\subsubsection{The case $p=2$}\label{sec-Dimp2}

As may be expected, in the case $p=2$, which corresponds to a (3+1)-dimensional universe (with the possibility of $q$ small dimensions compactified on a sphere), our results are similar to those of \cite{BoussoPhenom, BoussoPhenom1}. In Region I, corresponding to $t_{obs} < t_c < t_\Lambda$, the probability remains almost constant over $\log t_\Lambda$, but increases with decreasing $\log t_c$. In Region II, corresponding to $t_c < t_{obs} < t_\Lambda$, the probability remains almost constant over $\log t_\Lambda$, but increases with increasing $\log t_c$. In Region III, corresponding to $t_c < t_\Lambda < t_{obs}$, the probability increases with both $\log t_\Lambda$ and $\log t_c$. In Region IV, corresponding to $t_\Lambda < t_{obs}$, the probability remains constant over $\log t_c$ but increases with $\log t_\Lambda$, whereas in Region V, corresponding to $t_{obs} < t_\Lambda$, the probability remains constant over $\log t_c$ and decreases as $\log t_\Lambda$ increases.

This information allows us to draw the force diagram for the probability distribution over $\log t_c$ and $\log t_\Lambda$ for fixed $t_{obs}$, where we have assumed that the factor $g(\log t_c)\times \alpha(\log t_{obs},\log t_c, \log t_\Lambda)$ do \textit{not} provide the leading contribution to the probability of observations. We see that for any fixed $t_{obs}$, the maximum of the probability distribution lies along the lines $\log t_{obs} \sim \log t_\Lambda$ and  $\log t_{obs} \sim \log t_c$. By assuming that $g(\log t_c)$ decreases mildly, like an inverse power of $\log t_c$\cite{Weinberg}, and by making mild assumptions on the form of $\alpha$, as given in \cite{BoussoPhenom, BoussoPhenom1}, we can predict that the maximum of the probability distribution, when all three timescales are allowed to vary, is at
\begin{align}
\log t_{obs} \sim \log t_c \sim \log t_\Lambda \sim \log t_\Lambda^{max},
\end{align}
where $\log t_\Lambda^{max}$ corresponds to the smallest cosmological constant in the landscape, when considering bubbles with $p+1 = 3$ large dimensions. This scale is set by the number $\mathcal{N}$ of such vacua in the landscape, according to $t_\Lambda^{max} \sim \mathcal{N}^{1/2}$.

\begin{figure}\includegraphics[width=150pt]{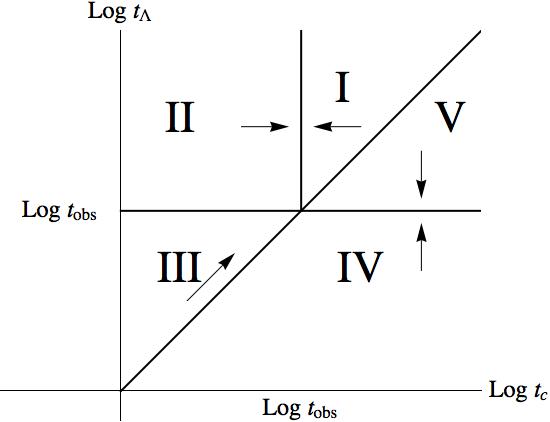}
\caption{The force diagram for the probability distribution when $p=2$ for fixed $t_{obs}$, obtained using the causal patch measure. The arrows indicate directions of increasing probability. The distribution is peaked along the degenerate half-lines forming the boundary between Regions I and II.}
\label{fig-Forcep2}
\end{figure}

%%%%%%%%
% 
%%%%%%%%%

\subsubsection{The case $p\geq 3$}\label{sec-Dimp3}

In the case $p \geq 3$, which corresponds to universes with a larger number of large dimensions than our own, we obtain slightly different results. In Region I, corresponding to $t_{obs} < t_c < t_\Lambda$, the probability increases with $\log t_\Lambda$, but decreases with $\log t_c$. In Region II, corresponding to $t_c < t_{obs} < t_\Lambda$, the probability increases with both $\log t_\Lambda$ and $\log t_c$. In Region III, corresponding to $t_c < t_\Lambda < t_{obs}$, the probability is sharply peaked at large $\log t_\Lambda$ and increases with $\log t_c$. In Regions IV and V, corresponding to $t_\Lambda < t_{obs}$ and $t_{obs} < t_\Lambda$ respectively, the probability remains constant over $\log t_c$ but increases with $\log t_\Lambda$.

Once again, we can draw a force diagram for the probability distribution over $\log t_c$ and $\log t_\Lambda$ for fixed $t_{obs}$. We see that for any fixed $t_{obs}$, there is a runaway of the probability distribution towards large values of $\log t_\Lambda$, and the distribution is peaked along the line $\log t_{obs} \sim \log t_c$. Thus once again, as long as the factors $g(\log t_c)\times \alpha(\log t_{obs},\log t_c, \log t_\Lambda)$ do not dominate the probability distribution, we can predict that the maximum probability, when all three timescales are allowed to vary, is at
\begin{align}
\log t_{obs} \sim \log t_c \sim \log t_\Lambda \sim \log t_\Lambda^{max},
\end{align}
where $\log t_\Lambda^{max}$ corresponds to the smallest cosmological constant in the landscape, when considering bubbles with $p+1$ large dimensions. So we see the same coincidence as in the case $p=2$, even though the probability distribution itself looks different. As before, we see that the scale $\log t_{obs}$ is set by the smallest cosmological constant in the landscape, when considering bubbles with $p+1 = 4$ large dimensions. This scale is set by the number $\mathcal{N}$ of such vacua in the landscape, according to $t_\Lambda^{max} \sim \mathcal{N}^{1/2}$.

\begin{figure}\includegraphics[width=150pt]{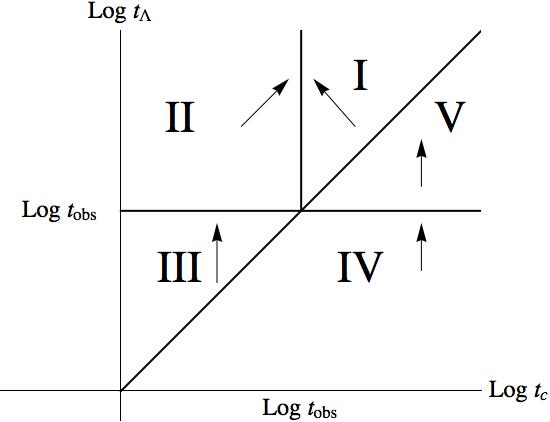}
\caption{The force diagram for the probability distribution when $p\geq 3$ for fixed $t_{obs}$, obtained using the causal patch measure. The arrows indicate directions of increasing probability. The distribution exhibits a runaway towards large $t_\Lambda$ and $t_c \sim t_{obs}$.}
\label{fig-Forcep3}
\end{figure}

\subsubsection{The case $p=1$}\label{sec-Dimp1}

Finally we consider the case $p=1$, which covers universes with one fewer large dimension than our own. In Regions I and II, corresponding to $t_{obs} < t_c < t_\Lambda$ and $t_c < t_{obs} < t_\Lambda$ respectively, the probability decreases as $\log t_\Lambda$ increases, but remains constant over $\log t_c$. In Region III, corresponding to $t_c < t_\Lambda < t_{obs}$, the probability decreases with $\log t_\Lambda$, seeming to reach a local maximum near the minimum value of $\log t_\Lambda$, and remains constant over $\log t_c$. In Region IV, corresponding to $t_\Lambda < t_{obs}$, the probability remains constant over $\log t_c$ but decreases with $\log t_\Lambda$, whereas in Region V, corresponding to $t_{obs} < t_\Lambda$, the probability remains constant over $\log t_c$ and increases as $\log t_\Lambda$ increases.

We can now draw the force diagram for the probability distribution over $\log t_c$ and $\log t_\Lambda$ for fixed $t_{obs}$. We see that for any fixed $t_{obs}$, there is a runaway of the probability distribution towards small values of $\log t_\Lambda$ in Regions I, II, and III, and the distribution is peaked along the line $\log t_{obs} \sim \log t_\Lambda$ in Regions IV and V. This situation is more complicated than in the cases $p\geq 2$, as we cannot make any definitive statements about coincidences of the timescales without more detailed calculations.

\begin{figure}\includegraphics[width=150pt]{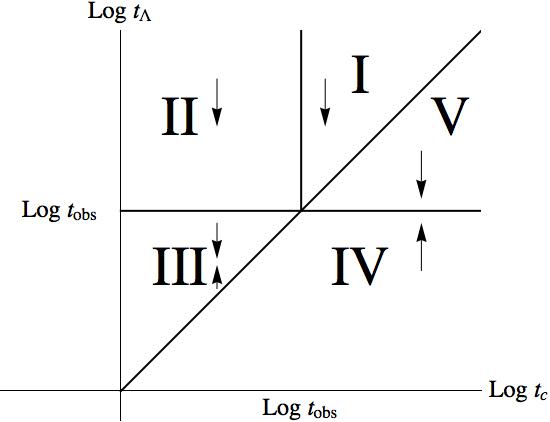}
\caption{The force diagram for the probability distribution obtainwhen $p=1$ for fixed $t_{obs}$, obtained using the causal patch measure. The arrows indicate directions of increasing probability. The distribution is peaked along $\log t_{obs} \sim \log t_\Lambda$ in Regions IV and V, and exhibits a runaway towards small $\log t_\Lambda$ in the rest of parameter space.}
\label{fig-Forcep1}
\end{figure}

\subsection{The fat geodesic measure}\label{sec-FG1}

For the fat geodesic measure, we can use the fact that we count observations in a fixed physical volume around the geodesic, and the solution for $a(t)$ found using the Friedmann equation, to determine the volume available to observers living at $t_{obs}$. Since we are counting observers in a fixed physical volume, the mass of observers within the cutoff region is proportional to the matter density:
\begin{align}
M_{FG} \alpha \rho_m \sim \frac{t_c^{p-1}}{a^{p+1}}.
\end{align}
The probability distribution of observations made in the landscape is then given by (\ref{eq-ProbDistCP}), but with $M_{FG}$, the mass of observers inside the region defined by the fat geodesic cutoff, replacing $M_{CP}$, the mass of observers inside the region defined by the causal patch cutoff.

As with the causal patch cutoff, we scanned over a range of $t_\Lambda$ by varying the parameters $p$, $q$, $Q$, and $\Lambda$, and solved the differential equations for $a(t)$ and $\phi(t)$ for a large range of $t_\Lambda$ and $t_c$ (where $t_c$ was input by hand), and then calculated $M_{FG}$ for a range of $t_{obs}$, choosing values for the three time scales so that they were well separated. We considered the same Regions $I--V$ of parameter space as with the causal patch measure. We then plotted the calculable parts of the probability distribution, namely $t_\Lambda^{-2}M_{FG}$, for each region and a variety of $(p,q,Q,\Lambda)$ to see if there were any discernible patterns in the results.

As with the causal patch measure, we found that the qualitative results depend only on the number $p+1$ of large dimensions. However, unlike the causal patch measure, in the case of the fat geodesic measure the shape of the probability distribution is the same in all cases where the number $p+1$ of large dimensions is $\geq 3$. These results are discussed in Sections \ref{sec-Dimp2Scale}-\ref{sec-Dimp1Scale}.

\subsubsection{The case $p \geq 2$}\label{sec-Dimp2Scale}

In the case $p \geq 2$, our results are similar to those of \cite{BoussoPhenom, BoussoPhenom1} in the case $p=2$. In Region I, corresponding to $t_{obs} < t_c < t_\Lambda$, the probability remains constant over $\log t_c$, but decreases with $\log t_\Lambda$. In Region II, corresponding to $t_c < t_{obs} < t_\Lambda$, the probability decreases with $\log t_\Lambda$, but increases with $\log t_c$. In Region III, corresponding to $t_c < t_\Lambda < t_{obs}$, the probability increases with both $\log t_\Lambda$ and $\log t_c$. In Region IV, corresponding to $t_\Lambda < t_{obs}$, the probability remains constant over $\log t_c$ but increases with $\log t_\Lambda$, whereas in Region V, corresponding to $t_{obs} < t_\Lambda$, the probability remains constant over $\log t_c$ and decreases as $\log t_\Lambda$ increases.

This information allows us to draw the force diagram for the probability distribution over $\log t_c$ and $\log t_\Lambda$ for fixed $t_{obs}$, where we have assumed that the factor $g(\log t_c)\times \alpha(\log t_{obs},\log t_c, \log t_\Lambda)$ do \textit{not} provide the leading contribution to the probability of observations. We see that for any fixed $t_{obs}$, the maximum of the probability distribution lies along the line separating Regions IV and V. A mild assumption on the prior probability distribution that favors small $\log t_c$\cite{BoussoPhenom, BoussoPhenom1} thus predicts the coincidence
\begin{align}
\log t_{obs} \sim \log t_c \sim \log t_\Lambda.
\end{align}
We have confirmed numerically that $M_{FG}$ decreases with $t_{obs}$. Assuming that the factor $g(\log t_{obs})$ in (\ref{eq-ProbDistCP}) also decreases with $t_{obs}$, and that $\alpha(\log t_{obs})$ does not grow too quickly with $t_{obs}$, we can say that there is a strong preference for $t_{obs}$, and thus $t_{\Lambda}$, to be small.

\begin{figure}\includegraphics[width=150pt]{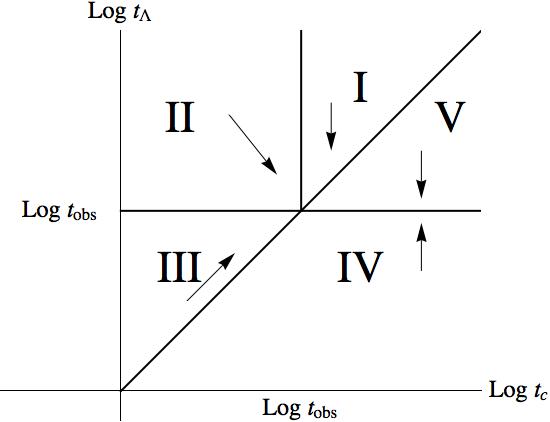}
\caption{The force diagram for the probability distribution when $p \geq 2$ for fixed $t_{obs}$, obtained using the fat geodesic cutoff. The arrows indicate directions of increasing probability. The distribution is peaked along the line separating Regions IV and V.}
\label{fig-Forcep2Scale}
\end{figure}

\subsubsection{The case $p=1$}\label{sec-Dimp1Scale}

Finally we consider the case $p=1$, which covers universes with one fewer large dimension than our own. In all the Regions $I--V$, the probability distribution is constant over $\log t_c$. In Regions $I--III$ and Region V, the probability decreases with $\log t_\Lambda$, whereas in Region $IV$ the probability increases with $\log t_\Lambda$ to reach a local maximum at some large value of $\log t_\Lambda$ at the upper limit of its range, such that $t_\Lambda < t_{obs}$.

We can now draw the force diagram for the probability distribution over $\log t_c$ and $\log t_\Lambda$ for fixed $t_{obs}$. We see that for any fixed $t_{obs}$, there is a runaway of the probability distribution towards small values of $\log t_\Lambda$ in Regions I, II, III, and V. In Region IV, the probability distribution increases with increasing $\log t_\Lambda$ towards a local maximum near the top of the range of $t_\Lambda$, when $t_\Lambda < t_{obs}$ and $t_\Lambda \lll t_c$. This situation is more complicated than in the cases $p\geq 2$, as we cannot make any definitive statements about coincidences of the timescales without more detailed calculations and better knowledge of prior probabilities.

\begin{figure}\includegraphics[width=150pt]{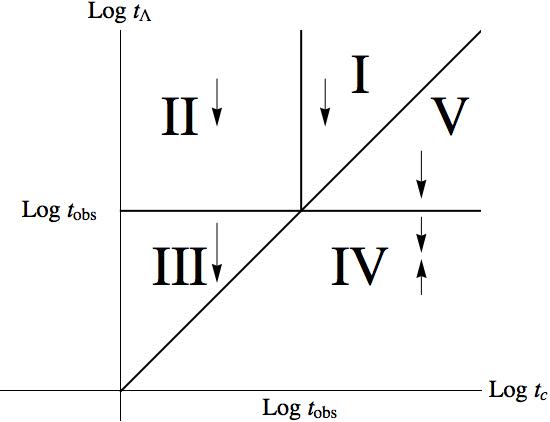}
\caption{The force diagram for the probability distribution obtained using the fat geodesic cutoff when $p=1$ for fixed $t_{obs}$, obtained using the fat geodesic cutoff. The arrows indicate directions of increasing probability. The distribution is peaked along $\log t_{obs} \sim \log t_\Lambda$ in Regions IV and V, and exhibits a runaway towards small $\log t_\Lambda$ in the rest of parameter space.}
\label{fig-Forcep1Scale}
\end{figure}

\section{Discussion}\label{sec-Conclusion}

In conclusion, we find that the prediction of the double coincidence of timescales $t_{obs} \sim t_{\Lambda} \sim t_c$, and the relation of the smallness of the cosmological constant to the size of the landscape, holds even when the multiverse is generalized to be multidimensional, both for the causal patch measure and the fat geodesic cutoff. Thus, the observed coincidence cannot be used to preferentially select one measure over the other when we are considering a multidimensional multiverse. Furthermore, this prediction holds regardless of the number $p+1$ of large dimensions in a given vacuum, which indicates that this observable cannot be interpreted as a consequence of our particular universe having three large spatial dimensions. In the case of the causal patch measure, the coincidence of timescales occurs for a different reason in the case when $p=2$ compared to when $p\geq 3$, as the probability distribution of observations has very different properties in the two cases. When the fat geodesic cutoff is generalized to a multi-dimensional multiverse, we find that the measure is even more robust: in addition to predicting the coincidence of the timescales $t_{obs} \sim t_{\Lambda} \sim t_c$ in all cases when the parameter $p \geq 2$, the shape of the probability distribution of observations is independent of the value of $p$. This makes intuitive sense when you consider that the generalization of the fat geodesic cutoff gives the same weighting to any unit of proper volume within a multiverse, regardless of the number of large dimensions. Thus, the probability distribution of observations remains independent of the number of large dimensions. We are thus led towards the conclusion that the coincidence of the timescales that we observe within our own universe is not a special feature of the particular features of our world.

There are many possibilities for further investigations along these lines. For example, although we have argued that the multiverse model we have considered is fairly generic, it could be worthwhile to investigate other models of multiverses with multi-dimensional vacua to see if the same patterns emerge, to make sure that these results are not model-dependent. It would also be profitable to carry out a more detailed analysis of the probability distribution in this model, to see if more distinctive features of the distribution could be uncovered in addition to identifying the maxima. It would be interesting to see, for instance, whether the agreement between $\log t_{obs}$, $\log t_\Lambda$, and $\log t_c$ varies with $p$, or whether the distribution becomes more sharply peaked along the lines of coincidence as $p$ varies. In general, the qualitative statement that the probability distribution of observations appears to change significantly when $p > 2$ compared to $p=2$ suggests that it would be worthwhile to extent the study of measures from multiverses with only (3+1)-dimensional vacua to multiverses with vacua of many different dimensions.

\section{Acknowledgments}

I would like to thank Professor Yasunori Nomura for helpful discussions on this subject and the measure problem in general, and for his comments on the draft of the manuscript.


\begin{thebibliography}{21}

\bibitem{Guth} A. H. Guth, ÒEternal inflation and its implications,Ó J. Phys. A \textbf{A40}, 6811-6826 (2007), arXiv:hep-th/0702178.

\bibitem{FreivogelReview} B.~Freivogel, \textit{Making predictions in the multiverse}, arXiv:1105.0244 [hep-th].

\bibitem{BoussoBoundary} R.~Bousso, B.~Freivogel, S.~Leichenauer, V.~Rosenhaus, \textit{Boundary definition of a multiverse measure}, Phys. Rev. \textbf{D82}, 125032 (2010), arXiv:1005.2783 [hep-th].

\bibitem{Coleman} S.~R.~Coleman, F.~De Luccia, \textit{Gravitational Effects on and of Vacuum Decay}, Phys. Rev. \textbf{D21}, 3305 (1980).

\bibitem{GarrigaProb} J.~Garriga, D.~Schwartz-Perlov, A.~Vilenkin, S.~Winitzki, \textit{Probabilities in the inflationary multiverse}, JCAP 0601, 017 (2006), arXiv:hep-th/0509184 [hep-th].

\bibitem{ScaleFactorCC} A.~De Simone, A.~H.~Guth, M.~P.~Salem, A.~Vilenkin, \textit{Predicting the cosmological constant with the scale-factor cutoff measure}, Phys. Rev. \textbf{D78}, 063520 (2008), arXiv:0805.2173 [hep-th].

\bibitem{ScaleFactor} R.~Bousso, B.~Freivogel, I-S.~Yang, \textit{Properties of the scale factor measure}, Phys. Rev. \textbf{D79}, 063513 (2009), arXiv:0808.3770 [hep-th].

\bibitem{Feldstein} B.~Feldstein, L.~J.~Hall, T.~Watari, \textit{Density perturbations and the cosmological constant from inflationary landscapes}, Phys. Rev. \textbf{D72}, 123506 (2005), arXiv:hep-th/0506235.

\bibitem{QCatastrophe} J.~Garriga, A.~Vilenkin, \textit{Anthropic prediction for Lambda and the Q catastrophe}, Prog. Theor. Phys. Suppl. 163, 245-257 (2006), hep-th/0508005.

\bibitem{Graesser} M.~L.~Graesser, M.~P.~Salem, \textit{The scale of gravity and the cosmological constant within a landscape}, Phys. Rev. \textbf{D76}, 043506 (2007), arXiv:astro-ph/0611694.

\bibitem{GarrigaPrediction} J.~Garriga, A.~Vilenkin, \textit{Prediction and explanation in the multiverse}, Phys. Rev. \textbf{D77}, 043526 (2008), arXiv:0711.2559 [hep-th].

\bibitem{ProbBoussoPolchinski} D.~Schwartz-Perlov, A.~Vilenkin, \textit{Probabilities in the Bousso-Polchinski multiverse}, JCAP 0606, 010 (2006), arXiv:hep-th/0601162.

\bibitem{FreivogelObsvConsq} B.~Freivogel, M.~Kleban, M.~Rodriguez Martinez and L.~Susskind, \textit{Observational consequences of a landscape}, JHEP 0603, 039 (2006), arXiv:hep-th/0505232.

\bibitem{PropTime} R.~Bousso, B.~Freivogel, I-S.~Yang, \textit{Boltzmann babies in the proper time measure}, Phys. Rev. \textbf{D77}, 103514 (2008), arXiv:0712.3324 [hep-th].

\bibitem{Vilenkin} A.~Vilenkin, \textit{Making predictions in an eternally inflating universe}, Phys. Rev. \textbf{D52}, 3365-3374 (1995), arXiv:gr-qc/9505031.

\bibitem{CPBousso} R.~Bousso, B.~Freivogel, I-S.~Yang, \textit{Eternal Inflation: The Inside Story}, Phys. Rev. \textbf{D74}, 103516 (2006), arXiv:hep-th/0606114.

\bibitem{CPBousso1} R.~Bousso, \textit{Holographic probabilities in eternal inflation}, Phys. Rev. Lett. 97, 191302 (2006). arXiv:hep-th/0605263.

\bibitem{Salem} M.~P.~Salem, \textit{Negative vacuum energy densities and the causal diamond measure}, Phys. Rev. \textbf{D80}, 023502 (2009), arXiv:0902.4485 [hep-th].

\bibitem{CAHPlus} M.~P.~Salem, A.~Vilenkin, \textit{Phenomenology of the CAH+ Measure}, Phys. Rev. \textbf{D84}, 123520 (2011), arXiv:1107.4639 [hep-th].

\bibitem{Weinberg} S.~Weinberg, \textit{Anthropic bound on the cosmological constant}, Phys. Rev. Lett. \textbf{59} (1987) 2607.
\bibitem{Giddings} S.~B.~Giddings, \textit{The fate of four dimensions}, arXiv:hep-th/0303031
\bibitem{BoussoPhenom} R.~Bousso, B.~Freivogel, S.~Leichenauer, and V.~Rosenhaus, \textit{Geometric origin of coincidences and hierarchies in the landscape}, arXiv:1012.2869
\bibitem{Vilenkin2} D.~Schwartz-Perlov and A.~Vilenkin, \textit{Measures for a Transdimensional Multiverse}, arXiv:1004.4567
\bibitem{Nomura1} Y.~Nomura, \textit{Physical Theories, Eternal Inflation, and Quantum Universe}, arXiv:1104.2324
\bibitem{Nomura2} Y.~Nomura, \textit{Quantum Mechanics, Spacetime Locality, and Gravity}, arXiv:1110.4630
\bibitem{Randall} S.~M.~Carroll, M.~C.~Johnson, and L.~Randall, \textit{Dynamical compactification from de Sitter space}, arXiv:0904.3115.
\bibitem{Vilenkin1} J.~J.~Blanco-Pillado, D.~Schwartz-Perlov, and A.~Vilenkin, \textit{Quantum Tunneling in Flux Compactifications}, arXiv:0904.3106
\bibitem{BoussoPhenom1} R.~Bousso, B.~Freivogel, S.~Leichenauer, and V.~Rosenhaus, \textit{A geometric solution of the coincidence problem, and the size of the landscape as the origin of hierarchy}, arXiv:1011.0714
\bibitem{Weinberg} S.~Weinberg, \textit{Anthropic bound on the cosmological constant}, Phys. Rev. Lett. 59 (1987) 2607.
\end{thebibliography}
\end{document}